
\magnification = 1200
\hsize= 17 truecm\vsize=25 truecm

\def\thf{\baselineskip=\normalbaselineskip\multiply\baselineskip
by 9\divide\baselineskip by 8}
\thf

\def\spose#1{\hbox to 0pt{#1\hss}}
\def\lta{\mathrel{\spose{\lower 3pt\hbox{$\mathchar"218$}}
     \raise 2.0pt\hbox{$\mathchar"13C$}}}
\def\gta{\mathrel{\spose{\lower 3pt\hbox{$\mathchar"218$}}
     \raise 2.0pt\hbox{$\mathchar"13E$}}}
\def\calk{ {\kappa} }
\def\Z{ Z }

\centerline{\it Contribution to the Session on (non baryonic) Dark Matter
in Cosmology,}
\centerline{\it XXXth Rencontres de Moriond, Villars-sur-Ollon,
Switzerland, January, 1995.}
\centerline{\it (ed. B. Guideroni \& J. Tran Thanh Van, Editions Fronti\`eres,
Gif-sur Yvette, 1995)}

\vskip 3 cm

\centerline{\bf HOW ABOUT VORTONS ?}

           \vskip 1 cm

\centerline{\bf B. Carter }
\medskip
\centerline{D\'epartement d'Astrophysique Relativiste et de Cosmologie, }
\smallskip\centerline{Observatoire de Paris, 92 Meudon, France. }

\vskip 6 cm
\parindent = 1 cm

\noindent
{\bf Abstract. }
\medskip
It has been suggested that cosmological dark matter may include a population
of vortons (meaning small centrifugally stabilised cosmic string loops)
as an outcome of (non-standard) electroweak symmetry breaking. The
implications for this conjecture of recent theoretical progress on
superconducting string theory are discussed, particularly in relation to
problems of stability. It is tentatively concluded that if the underlying
field theory provides a carrier field with mass not too far below that of
the relevant Higgs mass (which may be a few hundred G.e.V.) electroweak
string formation may conceivably produce enough vortons (with an extended
mass distribution peaking perhaps at several hundred T.e.V.) to provide a
marginally significant dark matter contribution.

\parindent=1 cm

\vfill\eject
\bigskip\noindent
{\bf 1. How are they formed?}
\smallskip

Many of the most commonly considered theories of G.U.T. and (non standard)
electroweak symmetry breaking predict the formation of vacuum vortex
defects[1]. If it is of the ``local" or ``gauged" type, its energy
will be concentrated within a microscopic radius $r_x\approx
m_x^{-1}$ where (using units such that $\hbar=c=1$) $m_x$ is the Higgs mass
scale associated with the symmetry breaking, so that on a macroscopic scale
the vortex can be represented by a thin ``cosmic" string model.
(The present discussion is not concerned with ``global" vortices,
with logarithmically divergent energy, for which a thin string description
is less accurate but still useful.) When a finite loop of such a string is
produced (whether during the symmetry breaking phase transition, or by a
transconnection process later on) one expects it to start in a
randomly excited state, whose energy will then dissipate by background
drag and radiation. It will thus be damped down either until it reaches
a ``vorton" state, meaning a stationary or quasi stationary
configuration[2][3] in which the energy is minimised with respect to
any parameters that are effectively conserved during this process, or else
until entirely disappears.

The present contribution updates an early discussion of this topic
at previous ``Moriond astrophysics" meeting, in which the term ``cosmic
ring" was used as an alternative to ``vorton"[4][5][6][7] to express
the property of circularity that was naively taken for granted until it
was discovered[8][9] that non circular equilibrium states also exist,
at least in the strict string limit $r_x<<\ell$ where $\ell$ is the loop
circumference.

The possibility of vorton formation  originally first pointed out by
Davis and Shellard[2][3], who were the first to fully appreciate the
implications of Witten's revelation[10] of the likely existence of mechanisms
that can easily allow the occurrence of currents on cosmic strings.
Earlier work on cosmic strings assumed that the underlying vortices were
of the simpliest Nielsen-Olesen type which are describable on a
macroscopic scale by the longitudinally Lorentz invariant Goto-Nambu string
model[1]. The absence of internal structure in this degenerate model makes
the centrifugalsupport mechanism inoperative, and so excludes the
existence of any vorton states.

Before the importance of the centrifugal support mechanism was recognised at
last[2][3], several workers investigated the possibility of what were termed
``cosmic springs", meaning magnetostatically supported ring
configurations[11][12][13] that unlike vortons were postulated to be
non - revolving (strictly static as opposed to stationary). Their existence
depended on the supposition that the Witten current is electromagnetically
coupled, which is very plausible, but it was soon realised that they could
hardly be even ephemerally viable under natural conditions if the relevant
coupling constant has its usual very small value $e^2\simeq 1/137<<1$,
because taken by itself the magnetic force would be far too weak to balance
string tension except in very weakly curved configurations whose radius would
have to be far too large for it to be realistic to consider them as isolated
systems.It has however been shown by Peter[14] that although the magnetostatic
effect can be expected to cause at most a very small modification of the
radius of a circular vorton state (just slightly breaking the degeneracy that
allows the existence of arbitrarily ``bent" equilibrium states[8][9] for
neutral string loops), on the other hand electrostatic support (a
possibility that was overlooked in the earlier work) may be more important.

\vfill\eject
{\bigskip\noindent}
{\bf 2. How rapidly do they revolve?}
\smallskip

One reason why it was not immediately realised that the dominant effect of
Witten type currents in vacuum vortices would be mechanical rather
electromagnetic was the unavailability at that time of a suitable string
model for describing their macroscopic behavior.  Some of the first attempts
to evaluate electromagnetic effects simply continued to represent the
vortices as simple Goto Nambu strings in which the currents were postulated
to evolve as if the effect of their inertia on the motion of the string were
entirely non existent[15]. Most of the earliest attempts to take account of
such mechanical effects relied[16][17] on a model constructed as a
particular kind of linearised perturbation of the Goto-Nambu model that had
been proposed at the outset for use in the weak current limit by Witten
himself[10]. A minority of earlier workers took up the rather more canny
suggestion[18] that a conducting string model previously introduced in a
very different physical context by Nielsen[19] should be used.

The consideration that none of these special models could deal with the
predicted effect[10] of current saturation (sometimes referred to as
``quenching") motivated the development[20][21] of a more general
category of elastic string models in which current dependence of an
appropriately non-linear nature could be allowed for. This category
consists of models governed by 2-dimensional (string worldsheet)
action integral specified by a Lagrangian scalar ${\cal L}$ that is given
as some (generically non-linear) function of a surface momentum covector
$\tilde p_\mu$ that is constructed as the tangentially projected gauge
covariant derivative of a scalar field $\phi$. In the application
to Witten's[10] bosonic superconducting model, $\phi$ is to be interpreted
as the phase of the carrier condensate in the vortex. In any model of this
category there will be a corresponding dynamically conserved surface current,
$c^\mu={\calk}\tilde p^\mu$ with ${\calk}=-2d{\cal L}/d\tilde w>0$
where $\tilde w=\tilde p_\mu\tilde p^\mu$, and a corresponding surface
stress momentum energy density tensor, which will be expressible[7][8][9]
in terms of the fundamental tangential projection tensor $\eta^\mu_{\ \nu}$
of the worldsheet simply as $T^\mu_{\ \nu}= c^\mu \tilde p_\nu+
{\cal L}\eta^\mu_{\ \nu}$.

The non zero eigenvalues of $T^\mu_{\ \nu}$ are  the  Lagrangian ${\cal L}$
and its dynamical conjugate $\Lambda={\cal L}-{\calk} \tilde w$
whose negatives are identifiable with the rest frame energy density $U$
and the string tension $T$ according to the specification $U=-{\cal L}$,
$T=-\Lambda$, in the ``magnetic case", $\tilde w>0$, for which the current
is spacelike, and $U=-\Lambda$, $T=-{\cal L}$ in the ``electric" case,
$\tilde w<0$, for which the current is timelike. They will be functionally
related in such a way that their difference is given by
$U-T=\tilde{\calk}\vert\tilde w\vert$. The ensuing characteristic speeds
$c_{_{\rm E}}$ and $c_{_{\rm L}}$ of extrinsic ``wiggle" perturbations and
sound type longitudinal ``woggle" perturbations respectively will be
given [21] by $c_{_{\rm E}}^{\,2}={T/ U}$, $c_{_{\rm L}}^{\,2}=
-{dT/ dU}\ .$

The availability of this category of elastic string models makes it
possible[6][7] to derive many important properties of vortons even in
the absence of detailed knowledge of the appropriate equation of state. As a
special case of a more general study allowing for a general stationary
gravitational of electromagnetic background field[8] it was established
that in the absence of any background fields an equilibrium configuration
for any such string model must (as one might have guessed) be {\it straight}
(which is evidently impossible for a closed string loop) except in the
special {\it transcharacteristic} case characterised by $v^2=c_{_{\rm
E}}^2$, where $v$ is the longitudinal running velocity as determined by the
timelike surface energy  eigenvector that is aligned with $c^\mu$ in the
``electric" regime where the current is timelike, and that is orthogonal to
$c^\mu$ in the ``magnetic" regime where the current is spacelike (which
was the only possibility considered in the earliest work). This centrifugal
force balance condition evidently implies that a vorton must have a
revolution period givel by $\ell\sqrt{U/T}$ where $\ell$ is its
circumference. When the current is electromagnetically coupled,
the transcharacteristic running condition is modified to
become a subcharacteristic running condition that (assuming circularity,
which is presumably justified in this case) has been found by
Peter[14] to have a form expressible in terms of the electric current
intensity $I=e\sqrt{\vert c^\mu c_\mu\vert}$, using a suitably adjusted
definition of the effective radius $r_\sigma$ of the current distribution,
as $v^2=\big(T-I^2 {\rm ln}\{\ell/r_\sigma\}\big)/\big(U+I^2
{\rm ln}\{\ell/r_\sigma\}\big)$.

The specification of the running velocity $v$ determines the ratio of the
integral quantum numbers characterising the vorton, which can be defined to
be the phase winding number $N$ and the total particle number $\Z$ associated
with the current (so that when it is electromagnetically coupled the total
electric charge will be $Q=e\Z$), according[6][7] to the specification
$\Z/N=2\pi{\calk}v$ in the ``magnetic" case $\tilde w>0$ and
$\Z/N=2\pi{\calk}/v$ in the ``electric" case $\tilde w<0$. The
equilibrium condition will thus be expressible in a form that is valid for
either case by
$${\vert \Z\vert\over\vert N\vert}=2\pi{\calk}\sqrt{ {\Lambda+\tilde w
{\calk}^2 e^2{\rm ln}\{\ell/r_\sigma\}
\over {\cal L}-\tilde w{\calk}^2 e^2{\rm ln}\{\ell/r_\sigma\}}}\ .$$

The modification arising from a non zero electromagnetic coupling constant
$e\simeq 1/\sqrt {137}$ in the foregoing formula can be expected to be
very small in the ``magnetic" case, i.e. when the current is spacelike due
to the saturation limit $c^\mu c_\mu\lta m_\sigma^{\,2}$ due to the
``quenching" effect that was predicted by Witten[10] and confirmed by more
detailed investigations such as that of Babul, Piran, and Spergel[22],
independently of whether the current is electromagnetically coupled or
not, where $m_\sigma\approx 1/r_\sigma$ is the mass of the carrier field whose
condensation on the string supports the current, since it is usually assumed
that this mass is at most of the order of the Higgs mass, $m_\sigma^{\,2}\lta
m_x^{\,2}$, which itself determines the magnitude of the energy density and
tension, $U\approx T\approx m_x^{\,2}$ in this case. If the carrier
mass is relatively small, the corresponding running speed will have to be
relativistic, with Lorentz factor $\gamma=1/\sqrt{1-v^2}$ given by
$\gamma\gta m_x/m_\sigma$.

To describe the very different situation that arises in the less
familiar ``electric" case, it is helpful to consider the specific form of
the equation of state, as derived in the first numerical
investigation of the timelike current regime on a Witten vortex,
by Peter. The computed form of the Lagrangian
in the absence of electromagnetic coupling[23][24] has a form that
has recently been found to be expressible rather accurately by an
analytic formula given[25] in terms of a pair of constant mass parameters
$m$ and $m_\star$, and of a dimensionless normalisation constant
$\kappa_{_0}$ (which is typically of order unity), by
$ {\cal L}=$ $-m^2+\big(1-{\tilde\chi/ m_\star^{\,2}}
\big)^{-1}\tilde\chi/2$ with $ \tilde\chi=-\kappa_{_0}\tilde w\ ,$
which gives ${\calk}=\kappa_{_0}\big(1-\tilde\chi/m_\star^{\,2}\big)^{-2}$,
in which $m\approx m_x$ and $m_\star\approx m_\sigma$, so that one expects to
have $m_\star^{\,2}<16 m^2$. Subject to this condition, the allowed range of
the variable $\tilde\chi$ will have a lower limit $-m_\star^{\,2}/3<\tilde
\chi$
where $c_{_{\rm L}}^{\,2}\rightarrow 0$ in the ``magnetic" regime, and an
upper limit $\tilde\chi<2m^2m_\star^{\,2}/\big(2m^2+m_\star^{\,2}\big)$ where
$c_{_{\rm E}}^{\,2} \rightarrow 0$ in the ``electric" regime.  The limit
(attributable to ``quenching") in the magnetic regime imposes a
corresponding limit
${\vert \Z/ N\vert}
\gta 2\pi\kappa_{_0}$
on the values of the quantum numbers $\Z$ and $N$ that are compatible with a
vorton equilibrium state. When the ratio $m/m_\star$ is small as one
expects, the ``wiggle" speed $c_{_{\rm E}}$ will be not only relativistic
but also supersonic over most of the allowed range for $\chi$ but it will
become subsonic, $c_{_{\rm E}}<c_{_{\rm E}}$, when $\tilde\chi>
(4m^2-m_\star^{\,2})/ (4m^2+ 2 m_\star^{\,2})$, which is attained for large
values of the charge to winding number ratio,
${\vert \Z/N\vert}
\gta8\pi\kappa_{_0} \big( {m_x/ m_\sigma}\big)^5$.

When the current is electromagnetically coupled, the equation of state is
modified in a manner that still needs further investigation, but that has
been shown by Peter[26] to be negligible as far as the ``magnetic" case
is concerned, and that seems likely remain unimportant even in the electric
case so long as the charge density remains well within the limit
$\sqrt{\vert c^\mu c_\mu\vert}\lta m_\sigma/e^2$ set by the threshold beyond
which charge leakage is to be expected. However if $m_\sigma^{\,2}
/m_x^{\,2}\lta 2e\simeq 1/6$ such modifications might be expected
to become important when the charge density approaches the upper
limit $\vert c^\mu c_\mu\vert< 2m^2\big(1+2m^2/m_\star^{\,2}\big)^3$
that would be allowed in the absence of coupling.

{\bigskip\noindent}
{\bf 3. How small can they be?}
\smallskip

For the string description to be valid, the loop circumference
$\ell$ must substantially exceed the effective radius $r_\sigma\approx
1/m_\sigma$ of the current carrying sheat that surrounds the defect core
with smaller radius $r_x\approx 1/m_x$. In terms of this value
$\ell>>r_\sigma\gta r_x$, the mass $M$ and the maximum value
for the angular momentum $J$ of a stationary string
loop will be given[6][7] by $M=\ell(U+T)$ and $ J
\leq $ $\vert \Z N\vert=\ell^2\sqrt{UT}/2\pi\ $.

Previous quantitative estimates[2][3][4][5] were restricted to the
relativistically revolving case, which is the only possibility for vortons
of ``magnetic" type, i.e. those in which (whether or not there is
electromagnetic coupling) the current is spacelike, and for which the Lorentz
factor is bounded below by $\gamma\gta m_x/m_\sigma$. For such relativistic
(and therefore, by the preceeding analysis, supersonic) vortons, one obtains
$\sqrt{\vert \Z N\vert} \simeq$ $ \ell m/\sqrt{2\pi} \approx\ell m_x\ $,
and $M\simeq 2\ell m^2\approx 2m_x\sqrt{\vert\Z N\vert}$,
in which the integers $\Z$ and $N$ must be of comparable magnitude, $\vert
\Z /N\vert\simeq 2\pi{\calk}\approx 2\pi\kappa_{_0}$, and not too small in
order to satisfy  the requirement that $\ell/r_\sigma$ should be reasonably
large.

In the ``electric" case, for which (whether or not there is electromagnetic
coupling) the current is timelike, there is also the possibility of a
qualititively different and so far little studied kind of vorton state whose
revolution speed $v$ is small compared with that of light, and can even
fall below the soundspeed  $c_{_{\rm L}}$. This arises because although
the possibility of the tension tending to zero (a so called ``spring" limit)
has been excluded for reasonable parameter values in so far as spacelike
currents are concerned[24], it does occur (at least in the absence of
electromagnetic coupling) for timelike currents. In this low tension limit,
$T\rightarrow 0$, the formula[25] quoted above gives $U\rightarrow
2m^2(2m^2+m_\star^2)/m_\star^{\,2}$, from which one obtains
the asymptotic relations
$${\vert \Z\vert\over\vert N\vert} \sim 2\pi\kappa_{_0}
\big(2m^2/m_\star^{\,2}\big)^{5/2} \sqrt{ 2m^2+m_\star^{\,2}\over T}\ ,
\hskip 1 cm \vert \Z N \vert\sim {\ell^2 m\over2\pi m_\star}
\sqrt{2(2m^2+m_\star^{\,2})T}\ .$$
In this limit, $\vert \Z/N\vert>>1$ one thus obtains $\ell\sim \vert
\Z\vert (m_\star/m)^3/\sqrt{\kappa_{_0}(2m^2+m_\star^2)}$. This will be
consistent with the requirement $\ell>>r_\sigma$ provided $\vert
\Z\vert>>(m_x/m_\sigma)^4$, which (since $\vert N\vert $ cannot be less that
one) will automatically be the case if the subsonicity condition $\vert
\Z/ N\vert\gta(m_x/m_\sigma)^5$ is satisfied.   The relatively high
quantum numbers needed for such slowly revolving vortons will be
statistically harder to obtain than those needed for the more familiar
relativistic kind, but for reasons discussed in the next section, the
subsonically rotating ones may be the kind that is most durable.

\vfill\eject
\bigskip\noindent
{\bf 4. How long can they last?}
\smallskip

Having recognised that there is a natural process whereby vortons can be
formed as an end product of energy dissipation by a dynamically evolving
string loop, the next question one needs consider is that of their
durability, a subject that needs a lot more work before firm conclusions can
be drawn. In a systematic approach to this problem, the first thing to be
checked is whether the equilibrium states are dynamically stable within the
framework of the simple uncoupled string description described above. A
general analysis of this problem has recently established[27] that an
isolated circular string loop equilibrium state will always be dynamically
stable if its running speed $v=c_{_{\rm E}}$ is subsonic or transonic.
According to the preceeding analysis, subsonic vortons need
$\vert Z/ N\vert\gta(m_x/m_\sigma)^5$. For lower values of the charge
quantum number $\vert \Z\vert$, the vorton will fall in the supersonic range
($c_{_{\rm E}}>c_{_{\rm L}}$) in which it has been found[27] that some
unstable states occur. A closer examination[28] has however shown that
while these can only have an ephemeral existence (with
lifetimes rarely more than a few times the light crosssing time
for the loop), there are  plenty of other supersonically rotating
vorton states that will be dynamically stable.

After this confirmation that many dynamically stable vorton string states
are available, one of the next questions to be considered is the
possibility of ``secular" instability on much longer timescales,
particularly due to  electromagnetic coupling which, despite its relative
 weakness might have a significant cumulative effect in the long run.
A well known example of such ``secular" instability  is provided by rotating
perfect fluid neutron star models, which it is caused by
gravitational radiation reaction in modes whose
relative propagation speed is slower than that of the stellar rotation[29].
Although no analogous (relatively counterrotating but nevertheless forward
rotating) modes  exist in subsonically rotating states,
they do exist in the more easily obtainable supersonic vorton states.
In vortons whose current is not electromagnetically coupled and for which
only relevant radiation mechanism is gravitational, one might expect that
(as is the case in practice for neutron stars) the growth of such an
instability would be too slow for it to be astrophysically relevant -- at
least for lightweight (electroweak unification) strings if not for
heavyweight (G.U.T.) strings. However such instability might
rapidly eliminate all electromagnetically coupled vortons except or
the subpopulation in the subsonic range, so that after a cosmologically
the only survivors of the less massive supersonic type would be neutral.
(A contrary effect of electromagnetic coupling that might however be
relevant[11] is its tendency
to enhance stability by reducing the running velocity $v$ slightly below the
wiggle velocity $c_{_{\rm E}}$, which will tend to broaden the range of
subsonic states.

Another kind of instability mechanism that was discussed when the concept of
a vorton was first introduced[2][3][30], but that has not yet been the
subject of a careful mathematical investigation, is a quantum barrier
tunnelling process involving spontaneous emission of the carrier particles
(with mass $m_\sigma$).  Experience with nuclear physics shows that
timescales for such processes are extremely difficult to estimate because
they are exponentially dependent on quantities that are very sensitive to
details of the internal structure, so that the question of whether the
lifetime is microscopically short or cosmologically long can
easily be affected by quite small uncertainties in the parameters of the
underlying model. The crudest naive estimate for the decay timescale $\tau$
for a typical tunnelling process is to suppose that, in units of the
relevant dynamical timescale, presumably of
order of the Higgs timescale $m_x^{-1}$, it will be given by
the exponential of a number that is roughly the number of wavelengths
involved in crossing the barrier. For a spin down process involving
cancellation of linear momentum losses from opposite sides of the vorton, it
is reasonable to guess that the relevant barrier width is of the order of
the vorton diameter $\ell/\pi$ while the relevant wavelength will be that
determined by the carrier field $m_\sigma$. According to this simple line of
reasonning, the decay lifetime for a vorton of the more easily obainable
relativistic type will be given in very rough order of
magnitude by ${\rm ln}\{m_x\tau\}\approx m_\sigma\ell/\pi\approx \sqrt{\vert
\Z N\vert}m_\sigma/m_x$.

According to more intricate but (because it involves questionable
assumptions at intermediate stages) not obviously more convincing order of
magnitude estimation by Davis[30], the factor $m_\sigma/m_x$ on the right of
the naive formula of the preceeding paragraph should be replaced by
$(m_\sigma/m_x)^3$.  If this were correct then for small values of
$m_\sigma/m_x$ a vorton would decay much more rapidly than predicted by
preceeding estimate. However pending more sophisticated calculations
(whose results are likely to be highly dependent on detailed assumptions
about the parameters of the underlying field theory) to determine what is
really most appropriate, the Ockham principle suggests that the simpler
formula should perhaps be preferred. In any case if $m_\sigma$ is not much
smaller that $m_x$ the uncertainty about the numerical coefficient will be
just as important as the question of the extra factor  $(m_\sigma/m_x)^2$.
If $m_\sigma$ is nearly the same as $m_x$, the two naive estimates agree in
predicting that in order for the vorton to survive to the present day one
needs something like $\sqrt{\vert \Z N\vert}\gta 10^2$. However for
$m_\sigma/m_x \simeq 10^{-1}$ my naive provisional estimate of the
requirement for survival to the present day is just changed to $\sqrt{\vert
\Z N\vert}\gta 10^3$, whereas the much cited Davis estimate[3][30][31] would
require $\sqrt{\vert \Z N\vert}\gta 10^5$.

The feasibility of the quantum emission processes to which such estimates
apply is presumably dependent on the availablity of suitable perturbed
destination states of reduced energy for the vorton. However the classical
analysis (and experience with ordinary laboratory superfluidity) suggests
that, at least in the bosonic case, the existence of such reduced energy
destination states depends on the flow speed of the condensate being
supersonic. This makes it reasonable to conjecture that the
subsonically rotating vorton states in the ``electric regime",
would in fact be entirely stable for all practical cosmological purposes.
(They would in principle still be able to decay a coherent manner, but this
would require a simultaneous transition of nearly all the quanta involved,
so that the marginally sufficient proportionality factor $\sqrt{\vert
\Z N\vert}$ in the estimate for ${\rm ln}\{m_x\tau\}$ would presumably need
to be replaced a much larger safety factor, perhaps of the order of
$\vert \Z N\vert$).

\vfill\eject
\bigskip\noindent
{\bf 5. How many are to be expected?}
\smallskip

The essential preliminary to formation of a vorton is the creation of its
dynamical string loop precursor, either at the time of cosmic string
formation or by a subsequent string transconnection process. According to
the standard picture[31][32][33] of cosmic string formation by the Kibble
mechanism when the cosmological temperature $\Theta$ drops below the value
$\Theta_x\approx m_x$ determined by the relevant Higgs mass scale, the
ensuing dynamics will initially be controlled by the damping effect of the
ambient background, which will rapidly smooth out short wavlength structure.
However structure characterised by a curvature radius $R$ large compared
with the relevant damping scale $\tau$ say will be overdamped, and hence
will be able to survive over a Hubble timescale $t=H^{-1}$ where
$H\approx\sqrt G\, \Theta^2$ if and only if $R\gta\xi$ where $\xi \approx
\sqrt{\tau t}$. The time scale of the damping (due to drag by the thermal
background) will be given by $\tau^{-1}\approx\beta\Theta^3 m_x^{-2}$
with a dimensionless drag coefficient $\beta$ that depends on the underlying
field theory[34] but that is expected to be of order of unity, which implies
that the smoothing length scale will be given by $\xi\approx G^{-1/4}m_x
\Theta^{5/2}$. So long as the temperature remains above a critical value
$\Theta_\ast \approx\sqrt G\, m_x^2$ below which $\tau$ exceeds the Hubble
time (so that drag becomes negligible and radiation reaction takes over as
the dominant damping mechanism) any lengthscale $R\gta\xi$ will be
associated[33] with a corresponding number density $n\approx R^{-3}$ of
wiggly (but, due to the overdamping, not wriggly) string loops of
average length $L\approx R^2/ \xi$.

It is during this overdamping epoch characterised by $\sqrt G\, m_x^2\lta
\Theta\lta m_x$ that the condensation of the carrier field at a critical
temperature $\Theta_\sigma\approx m_\sigma$ is presumed to  occur. The
crucial step[2][3] in the formation of proto-vortons is a random walk type
process whereby one expects  that $\Z$ and $N$ will
acquire a Gaussian distribution with mean square value given by the
ratio of the loop circumference to the thermal fluctuation scale, i.e.
$\langle \Z^2\rangle\approx \langle N^2\rangle\approx m_\sigma L$. The
minimum initial loop circumference needed for producing  vortons
characterised by  conserved numbers at least the magnitude of given  values
$\vert\Z\vert \gta \vert N\vert $ with reasonable efficiency will thus be
given roughly by $L \gta\Z^2/m_\sigma$. Since the smoothing length
at this time will be given by  $\xi_x\approx G^{-1/4}m_x m_\sigma^{\,-5/2}$,
this means that to obtain $\Z^2\gta$  $(Gm_x^{\,2})^{-1/4}
(m_x/m_\sigma)^{3/2}$ one needs a corresponding minimum curvature length scale
 $R\gta$ $ G^{-1/8}m_x^{\,1/2}m_\sigma^{\,-7/4}{\vert \Z\vert}$.

If a fraction $f$ of such loops survives in the form of vortons at a later
epoch, their number density will be given in terms of this initial curvature
length scale $R$ by $n\approx\ f R^{-3}(\Theta/\Theta_\sigma)^3$ due to the
cosmological expansion as the universe cools from the carrier condensation
temperature $\Theta_\sigma$ to the lower temperature $\Theta$ at the later
epoch under consideration. The corresponding cosmological mass density
contribution will be $\rho\approx$ $Mn$ where $M$ represents the typical
individual mass energy value for such vortons, which will be given by
$M\approx$ $m_\sigma {\vert \Z\vert}$ for the supposedly more durable slowly
rotating kind needing a large value of $\vert\Z\vert$ and by $M\approx$
$m_\sigma {\vert Z\vert}$ for the initially more numerous relativistic kind,
so that for the latter one obtains $\rho\approx$ $f
G^{3/8}m_x^{\,-1/2}m_\sigma^{\,9/4} \Z^{-2}\Theta^3$.

It is of particular interest to evaluate the corresponding cosmological
closure fraction ${\mit\Omega}=\rho/\rho_{\rm c}$ as evaluated with
respect to the minimum mass density compatible with cosmological closure
(according to the standard Einstein theory in the absence of a cosmological
constant) which is given by $\rho_{\rm c}\approx m_{\rm c}\Theta^3$ where
$m_{\rm c}$ is the required mass energy per black body photon which is of the
order of $10^2\, e.V.$, i.e. $\sqrt{ G m_{\rm c}^{\,2}}\approx 10^{-26}$.
The preceeding formulae provide the estimate
$${\mit\Omega}\approx f \big(Gm_x^{\,2}\big)^{3/8} \Big({m_\sigma\over m_x}
\Big)^{9/4}Z^{-2}{m_x\over m_c}\ $$
for the contribution from vortons with $\vert Z\vert\gta \vert N\vert
\gta(Gm_x^{\,2})^{-1/8} (m_x/m_\sigma)^{3/4}$ if they
are of the relatistic kind, while if only the subsonic kind survives
the final factor $m_x/ m$ should be replaced by the rather smaller
factor $m_\sigma/m_c$, a change that is not very important in view of the
other uncertainties (such as the values of the numerical ``order of unity"
numerical factors $\kappa_{_0}$ and $\beta$). It can be seen that as $\vert
\Z\vert$ is increased above this smoothing limit the
vorton spectrum decreases rather steeply. On the other hand the spectrum
levels off for values below this limit so that the total
subsonic contribution for moderate and small quantum number values
will be given just by
${\mit\Omega}\approx  f \big(G m_x^{\,2}\big)^{5/8} \Big({m_\sigma/
m_x}\Big)^{11/4} { m_x/ m_c}\ $.
The typical mass of a vorton in this sub population near the lower cut off
limit (which would constitute the bulk of the population if $f$ remained
large for such low quantum number values) will be given by $M\approx
m_x(Gm_x^{\,2})^{1/8}(m_\sigma/m_x)^{1/4}$.

The enormous value $m_x/m_c\approx 10^{23}$ obtained for the final factor of
the preceeding formula in the G.U.T. case makes it hard to see how the
heavyweight string scenario can avoid excessive vorton production if there is
any carrier field with mass $m_\sigma$ that is even remotely comparable with
the Higgs mass. Even for lightweight strings, with $m_x/m_\sigma\approx
10^{10}$, the value obtained for ${\mit\Omega}$ would not be entirely
negligible compared with unity if $m_\sigma$ were nearly as large as $m_x$
and if the survival fraction $f$ is reasonably large. In order for $f$ to
avoid being negligibly small $\Z$ must be large enough to allow long term
survival, which would seem likely to require at least $\Z\gta 10^2$ for
avoidance of decay by quantum tunnelling in the rapidly revolving case, and
$Z\gta(m_x/m_\sigma)^5$ for stabilisation in a subsonically rotating state,
which is a more stringent requirement if $m_\sigma/m_x$ is very small. However
provided $m_\sigma/m_x\gta 10^{-1}$ even this requirement will be satisfied by
most of the loops in the distribution, which (in the electroweak case
characterised by $\sqrt{Gm_x^{\,2}}\approx 10^{-16}$) will peak at a
smoothing limit given $\Z\approx 10^4$

Most such vortons would be on the high side of the multi  T.e.V. mass range
that has been envisaged for hypothetical ultramassive dark matter particles
(known, if they are charged, as ``CHUMPs" or ``CHAMPS). The expectation that
their density would be too low for them to contribute much towards
cosmological closure means that the formation of electroweak vortons cannot
yet be ruled out by direct detection limits[35][36][37] or even by the
indirect  considerations[38] that place even more severe constraints on their
abundance unless they are of the electrically uncoupled kind.
The best chance of detecting a sparse but electrically coupled vorton
distribution might be as cosmic rays. In particular, it has been suggested
that rare events of extremely high energy that are hard to account for in
terms of the usual acceleration mechanisms [39] might plausibly be
attributed to vortons[40].

 \parindent= 0 cm
\bigskip
{\bf References}
\smallskip

[1] T.W.B. Kibble, {\it J.Phys.} {\bf A9}, 1387 (1976).

[2] R.L. Davis, E.P.S. Shellard, {\it Phys. Lett.} {\bf B209},
485 (1988).

[3] R.L. Davis \& E.P.S. Shellard, {\it Nucl. Phys.} {\bf B323}, 209 (1989).
\vfill\eject
[4] B. Carter, in {\it Early Universe and Cosmic Structures} (Xth Moriand
Astrophysics Meeting), ed J-M. Alimi, A. Blanchard, A. Bouquet, F. Martin de
Volnay \& J. Tran Than Van, 213 (Editions Fronti\`eres, Gif sur Yvette, 1990).


[5] B. Carter, {\it Ann. N.Y. Acad. Sci.} {\bf 647}, 758 (1991).

[6] B. Carter, {\it Phys. Lett.} {\bf B238}, 166 (1990).

[7] B. Carter, in {\it The Formation and Evolution of Cosmic Strings},
ed G. Gibbons, S. Hawking \& T. Vachaspati (Cambridge U.P., 1990).

[8] B. Carter, V.P. Frolov \& O. Heinrich, {\it Class. and Quantum Grav.}
{\bf 8}, 135 (1991).

[9] B. Carter, {\it Class. Quantum Grav.} {\bf 9S}, 19 (1992).

[10] E. Witten, {\it Nucl. Phys.} {\bf B249}, 557 (1985).

[11] E. Copeland, M. Hindmarsh \& N. Turok, {\it Phys. Rev. Lett.}
{\bf 58}, 1910 (1987)

[12] E. Copeland, D. Haws, M. Hindmarsh \& N. Turok, {\it Nucl.
Phys.} {\bf B306}, 908 (1988).

[13] D. Haws, M. Hindmarsh \& N. Turok, {\it Phys. Lett.} {\bf B209},
225 (1988).

[14] P. Peter, {\it Phys. Lett.} {\bf B298}, 60 (1993).

[15] D.N. Spergel, W.H. Press, R.J. Scherrer {\it Phys. Rev.}{\bf D39},
379 (1989).

[16] D.N. Spergel, T. Piran, J. Goodman, {\it Nucl. Phys.}
{\bf B291}, 847 (1987).

[17] A. Vilenkin, T. Vachaspati, {\it Phys. Rev. Lett.} {\bf 58},
1041 (1987)

[18] N.K. Nielsen, P. Olesen, {\it Nucl. Phys.} {\bf B291}, 829 (1987).

[19] N.K. Nielsen, {\it Nucl. Phys.} {\bf B167}, 248 (1980).

[20] B. Carter, {\it Phys. Lett.} {\bf B224}, 61 (1989).

[21] B. Carter, {\it Phys. Lett.} {\bf B228}, 446 (1989).

[22] A. Babul, T. Piran, D.N. Spergel, {\it Phys. Lett.}
{\bf B202}, 207 (1988).

[23] P. Peter, {\it Phys. Rev.} {\bf D45}, 1091 (1992).

[24] P. Peter, {\it Phys. Rev.} {\bf D47}, 3169 (1993).

[25] P. Peter, B. Carter, {\it DAMTP preprint R95159}; HEP-ph/9411425
(Cambridge, 1994).

[26] P. Peter, {\it Phys. Rev.} {\bf D46}, 3335 (1992).

[27] B. Carter\& X. Martin, {\it Ann. Phys.} {\bf 227}, 151 (1993).

[28] X. Martin, {\it Phys. Rev.} {\bf 50}, 7479 (1994).

[29] J.L. Friedman \& B.F. Schutz, {\it Astroph. J.} {\bf 222}, 881 (1978).

[30] R.L. Davis, {\it Phys. Rev.} {\bf D38}, 3722 (1988).

[31] A. Vilenkin \& E.P.S. Shellard, in {\it Cosmic Strings and other
Topological Defects}, 359 (Cambridge U.P., 1994).

[32] T.W.B. Kibble, {\it Physics Reports} {\bf 67}, 183 (1980).

[33]  T. Vachaspati \& A. Vilenkin, {\it Phys. Rev} {\bf D30}, 2036
(1984).

[34] A. Vilenkin, {\it Phys. Rev} {\bf D43}, 1046 (1991).

[35] A.S. De Rujula, S.L. Glashow \& U. Sarid, {\it Nucl. Phys.} {\bf B333},
173 (1990).

[36] J.L. Basdevant, R. Mochkovich, J. Rich, M. Spiro \& A. Vidal-Madjar,
{\it Phys. Lett.} {\bf B234}, 395 (1990).

[37] S. Dimopoulos, D. Eicher, R. Esmailzadeh \& G.D. Starkman, {\it Phys.
Rev.} {\bf D41}, 2388 (1990).

[38] A. Gould, B.T. Draine \& R.W. Romani, {\it Phys. Lett.} {\bf B238},
337 (1990).

[39] G. Sligl, D.N. Schramm \& P. Bhattacharrjee, {\it Astroparticle
Phys.} {\bf 2}, 401 (1994).

[40] P. Peter, in {\it Proc. Workshop on Giant Airshower Detectors}
(Fermilab, 1955).

 \end